\documentstyle[prd,aps,psfig]{revtex}
\def \beq{\begin{equation}}
\def \eeq{\end{equation}}
\def \beqa{\begin{eqnarray}}
\def \eeqa{\end{eqnarray}}
\def \alphae{\alpha_{\scriptscriptstyle E}}
\def \alphas{\alpha_{\scriptscriptstyle S}}
\def \alphav{\alpha_{\scriptscriptstyle V}}

\def \lms{\Lambda_{\overline{\scriptscriptstyle MS}}}
\begin{document}
\draft
\title{A precise determination of $T_c$ in QCD from scaling}
\author{Sourendu Gupta \footnote{Electronic mail: sgupta@tifr.res.in}}
\address{Department of Theoretical Physics, Tata Institute of Fundamental
         Research,\\ Homi Bhabha Road, Mumbai 400005, India.}
\maketitle
\begin{abstract}
Existing lattice data on the QCD phase transition are analyzed in
renormalized perturbation theory. In quenched QCD it is found that
$T_c$ scales for lattices with only 3 time slices, and that
$T_c/\lms=1.15\pm0.05$. A preliminary estimate in QCD with two flavours
of dynamical quarks shows that this ratio depends on the quark mass.
For realistic quark masses we estimate $T_c/\lms=0.49\pm0.02$.  We also
investigate the equation of state in quenched QCD at 1-loop order in
renormalised perturbation theory.
\end{abstract}
\pacs{11.15.Ha, 12.38.Mh\hfill TIFR/TH/00-56, hep-lat/0010011}

The QCD phase transition temperature ($T_c$) is a fundamental constant
of the hadronic world, and will soon be accessible to experiments. It
has also been the target of many lattice computations.  Current
practice is to express $T_c$ in units of the mass of the rho meson
($M_\rho$) or the square root of the string tension ($\sqrt\sigma$)
\cite{karsch}. However, in the last few years it has become clear that
the renormalised QCD coupling, $\alphas$, measured on fairly coarse
lattices \cite{lepage} yield results comparable to those obtained in
precision measurements at LEP and in other experiments \cite{pdg}. This
prompts us to test the approach to the continuum limit of QCD
thermodynamics by testing the constancy of $T_c/\lms$, where $\lms$ is
the QCD scale parameter extracted in the $\overline{\rm MS}$ scheme.

In the limit when all quark masses are zero (or infinite), QCD has only
one dimensionless parameter--- the coupling, $\alphas=g^2/4\pi$. Quantum
corrections transmute it into a momentum scale. This scale is either
given explicitly as the QCD parameter $\Lambda$, or specified implicitly,
as the value of the coupling $\alphas(\mu)$ at scale $\mu$.

In the lattice regularisation of QCD, the value of the lattice spacing
($a$) is determined by the bare coupling $\beta=6/g^2$. However,
$6/\beta$ is not a good expansion parameter for perturbation theory.
It is more appropriate to define the coupling through some physically
motivated process. In one definition, called the V-scheme
\cite{lepage}, $\alphas$ at scale $3.4018/a$ is found from the
logarithm of the plaquette value ($\cal P$) \footnote{ $\cal P$ is defined
to be one third of the real part of the trace of the product of four
link matrices taken in order around a plaquette.} through the formula
\beq
   -\ln{\cal P} = \frac{4\pi}3\alphav\left(1-(1.1897+0.071 N_f)\alphav\right).
\label{def}\eeq
In another definition, the E-scheme \cite{echeme}, the renormalized coupling
at scale $a$ is
\beq
   \alphae = \frac3{4\pi}\biggl(1-{\cal P}\biggr),
\label{defe}\eeq
We choose to work in these two schemes, as well as the $\overline{\rm MS}$
scheme whose relation to the V-scheme has been worked out \cite{brodsky}.
At 2-loop order we can write
\beq
   a\Lambda = k R\left(1/4\pi\beta_0\alphav\right), \qquad{\rm where}\qquad
   R^2(x)=\exp(-x)x^{\beta_1/\beta_0^2}. 
\label{scal}\eeq
The constant $k$ depends on the scheme, being unity in the E-scheme and
3.4018 in the V-scheme. The function $R$ is obtained by integrating the
two-loop beta function, $\overline\beta(g)=-\beta_0g^3-\beta_1g^5$.

The QCD phase transition temperature is determined by tuning the bare
coupling on lattices with $N_t\ll N_s$ (where $N_t$ is the number of
sites in the Euclidean time direction and $N_s$ that in the spatial
directions). Then, $T_c=1/a_cN_t$, where $a_c$ is the lattice spacing
at the coupling where the phase transition occurs. Our strategy is to
compute the renormalised coupling $\alphav$ at these bare couplings and
hence determine $T/\lms$ using eq.\ (\ref{scal}), and the known ratio
of $\lms$ and the scale $\Lambda$ in any other scheme.

In quenched QCD with the usual Wilson action, the critical bare couplings,
$\beta_c$, have been determined for $2\le N_t\le16$ \cite{ukawa}. The
main source of systematic uncertainty in the older data arises from the
fact that the thermodynamic limit $N_s\to\infty$ was not taken. Later
data \cite{iwasaki,boyd,jochen} have taken this limit, and we only used these
to study scaling. The statistical errors in these later studies are also
much smaller, and hence they are able to test the scaling hypothesis
much more stringently.

\begin{figure}[htbp]\begin{center}
   \leavevmode
   \psfig{figure=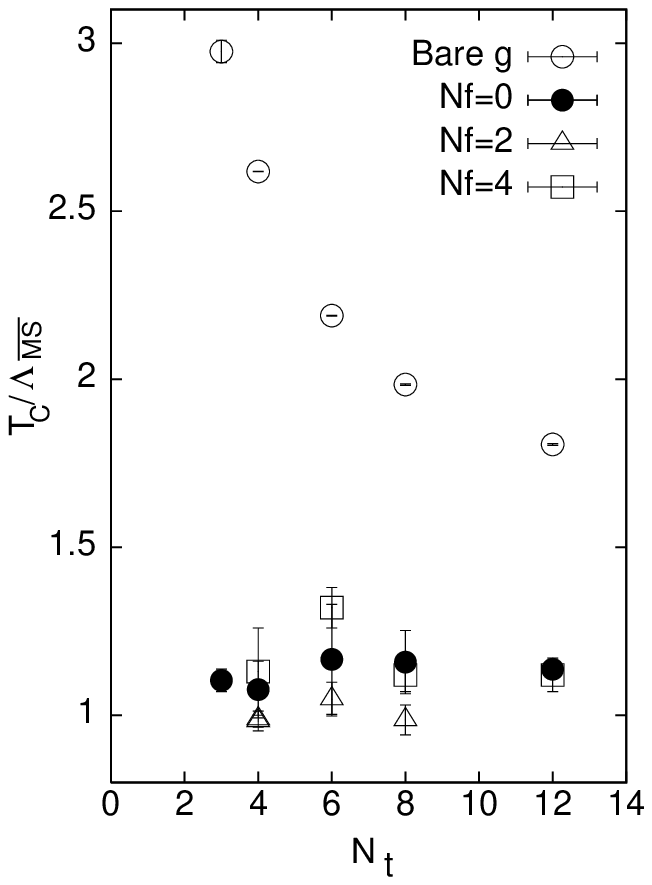,height=8cm,width=5.3cm}
   \psfig{figure=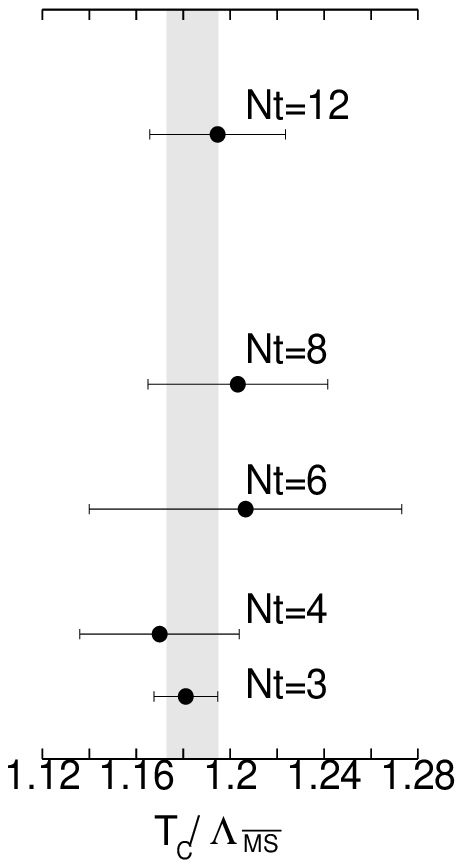,height=8cm,width=2.4cm}
   \end{center}
   \caption{The panel on the left shows $T_c/\lms$ as a function of
       $N_t$ in the V-scheme. For $N_f=0$ we show tests of scaling for
       both the bare lattice coupling and $\alphav$. For the former
       the errors come only from the determination of the critical
       coupling, whereas for the latter they also include errors in
       the determination of $\cal P$. For $N_f=2$ with $m/T_c=0.1$
       and $N_f=4$ with $m/T_c=0.08$ only the latter is shown. The
       panel on the right shows details of $T_c/\lms$ for $N_f=0$ in
       the $\overline{\rm MS}$-scheme.}
\label{fg.tcms}\end{figure}

We extracted $\alphas$ from the plaquette values listed in
\cite{jochen} using eqns.\ (\ref{def},\ref{defe}).  Values of $\ln{\cal
P}$ at $\beta_c$ were obtained by cubic spline interpolation.
Statistical errors in the interpolated values were found by
propagation.  We probed systematic errors in the interpolation by the
change in $\alphas$ on removal of some of the knot points. The results
of our analysis are shown in Figure \ref{fg.tcms}.  $T_c/\lms$ is
constant down to $\beta_c$ for $N_t=3$.  A large part of the error in
this ratio comes from the errors in the measurement of $\cal P$, which,
while small, are exponentiated in $\Lambda$.

More detailed results are shown in the right hand panel of Figure
\ref{fg.tcms}. Small scaling violations at these couplings, seen in
measurements at $T=0$, have been attributed to finite lattice spacing
errors \cite{allton}.  On replacing the scaling function $R$ in eq.\
(\ref{scal}) by $\overline{R}(\alphav) = R(\alphav)\bigl[ 1 +
c_2\hat{a}^2 + c_4\hat{a}^4 \bigr]$, where $\hat
a=R(\alphav)/R(\alphav^0)$ and $\alphav^0$ is determined at $\beta=6$,
it was found \cite{heller} that data on $a\sqrt\sigma$ could be
quantitatively described down to $\beta=5.4$.  As shown in the figure,
in the $\overline{\rm MS}$-scheme we can fit a constant value of
$T_c/\lms$ using the critical couplings determined for $3\le N_t\le14$
with good confidence--- $\chi^2=0.6$ for 4 degrees of freedom. The
statistical error in this estimate is about 1\%.  Taking into account
the variation between values obtained in different schemes, and for
fits with $N_t\ge3$ or 4, we quote
\beq
   T_c = (1.15\pm0.05) \lms\qquad\qquad\qquad\qquad({\rm Quenched\ QCD}).
\label{tclms}\eeq
This error estimate now includes not only the statistical errors but
also the systematic uncertainties above.  Since experiments cannot
access a quenched theory, the value of $\lms$ in MeV units is hard to
pin down. When comparing different estimates, in MeV units, of the same
scale in quenched QCD this universal problem should be borne in mind.
Here we convert $T_c$ into physical units using the value
$\lms\approx250$ MeV as a mean of the values obtained in different
schemes in \cite{heller} by assuming $\sqrt\sigma=465$ MeV. This gives
$T_c\approx285\pm10$ GeV, in reasonable agreement with other estimates,
such as that found in a recent study using an RG-improved action
\cite{kanaya}.

This analysis leads us to investigate the extraction of thermodynamic
quantities in the QCD plasma using renormalised perturbation theory.
Failure of scaling would then be a direct signal for lattice artifacts
such as power corrections in $a$.  We examine the energy density ($E$)
and the pressure ($P$). These can be written in terms of the difference
$\Delta_i={\cal P}_i-{\cal P}_0$ between the spatial ($i=s$) and
temporal ($i=t$) plaquettes and their zero temperature counterpart
${\cal P}_0$. $E$ is defined by the formula
\beq
   \frac E{T^4} = 6N_cN_t^4\left[\frac{\Delta_s-\Delta_t}{4\pi\alphas}
        -(c_s'\Delta_s+c_t'\Delta_t)\right].
\label{ene}\eeq
The anisotropy coefficients $c_s'$ and $c_t'$ are known to 1-loop order
\footnote{At this order the coefficients remain unchanged in going from
the lattice scheme, $g^2=6/\beta$, in which they have been computed, to
the $\overline{\rm MS}$ or V-scheme, but change in the E-scheme. In the
absence of a two-loop computation, we have evaluated the renormalised
couplings at the scale appropriate to the plaquette.}
\cite{karschcoff}.  A measure of deviations from ideal gas behaviour is
\beq
   \Delta = \frac{E-3P}{T^4} = 12N_cN_t^4(c_s'+c_t')(\Delta_t+\Delta_s),
\label{emp}\eeq
which can be combined with eq.\ (\ref{ene}) to give the pressure.  The
sum rule $g^3(c_s'+c_t')=\overline\beta(g)$ \cite{karschcoff}, allows
us to evaluate $\Delta$ beyond 1-loop order. In fact, part of the
finite $a$ corrections can be incorporated into $\Delta$ by evaluating
$\overline\beta(g)$ using $\overline R$ instead of $R$. However, we
avoid this approach, since we need to use $c_s'$ and $c_t'$
consistently in eqs.\ (\ref{ene},\ref{emp}). In the following we use
only the 1-loop results for these coefficients and work in the
$\overline{\rm MS}$ scheme.

\begin{figure}[htbp]\begin{center}
   \leavevmode
   \psfig{figure=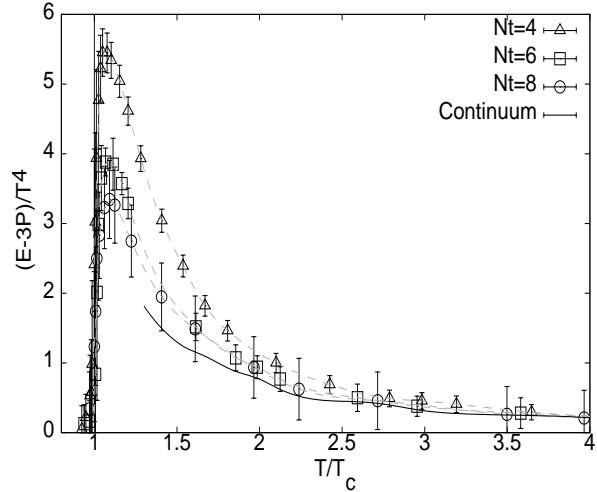,height=6.5cm,width=8cm}
   \end{center}
   \caption{$(E-3P)/T^4$ for a QCD plasma evaluated as a function of
       $T/T_c$ at different lattice spacings when
       $\alpha_{\overline{\scriptscriptstyle MS}}$ sets the scale. The
       continuum curve is not continued into the region where finite
       volume effects may be important. Errors on the continuum extrapolation
       are roughly similiar to those at $N_t=8$.}
\label{fg.thermo}\end{figure}

Raw data on ${\cal P}_{0,s,t}$ at a range of couplings measured on
$N_t\times(4N_t)^3$ lattices ($N_t=4$, 6 and 8) are tabulated in
\cite{boyd}. We have used these to evaluate thermodynamic quantities
only for $T<4T_c$, because the finite spatial volumes of the lattices
used may cause spatial deconfinement above $4T_c$. As expected,
$\Delta$ varies as $1/N_t^2$ at fixed $T$, {\sl i.e.\/}, as $a^2$. This
power correction can be removed through the use of improved actions
\cite{beinlich}.  $\Delta$ shows a peak at $T\approx1.1T_c$ in the
continuum limit.  However, the location of the peak is uncertain
because there is no unique definition of $T_c$ at finite volumes.
Different definitions, which all coincide in the thermodynamic limit,
give different values of the pseudo-critical point on finite volume
systems \cite{fse}. Using coarse lattices, we have estimated that in
$SU(3)$ theory for $N_s=4N_t$ this inherent uncertainty in the critical
coupling may be as much as $\delta\beta\approx0.005$--$0.01$: much
larger than the statistical error for any given definition of the
critical point. For similar reasons the value of $\Delta$ at the peak
cannot be reliably extracted without taking into account finite volume
effects.  In the range $1.3T_c\le T\le 4T_c$, finite volume corrections
are expected to be small. For such $T$, $\Delta$ is monotonically
decreasing. However, many quantities of interest can be extracted only
near $T_c$. The results of a finite size scaling study which does this
will be reported elsewhere.

We have also evaluated $E/T^4$ and found that it scales to the
continuum limit as $a^2=1/N_t^2$. In the temperature range where finite
volume effects are strong, the large value of $\Delta$ gives a negative
value to $P$ evaluated using the formulae in
eqs.\ (\ref{ene},\ref{emp}). As we have argued already, an evaluation
of thermodynamics in this range of $T$ requires better control over
finite volume effects. However, the problem of negative pressures is
also avoided if the continuum limit of $P$ is found by combining
$E/T^4$ and $\Delta$ in the limit $a=0$. This gives nearly vanishing
$P$ in the region of the peak in $\Delta$.

\begin{figure}[bhtp]\begin{center}
   \leavevmode
   \psfig{figure=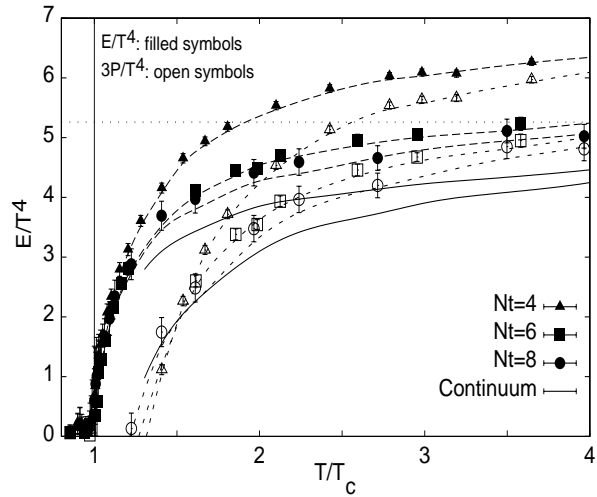,height=6.5cm,width=8cm}
   \end{center}
   \caption{$E/T^4$ and $3P/T^4$ for a QCD plasma evaluated as explained
       in the text, using the data of \protect\cite{boyd}. The dashed
       curves are smooth cubic spline fits. The dotted horizontal line
       is the ideal gas result in the continuum limit.}
\label{fg.cured}\end{figure}

In Figure \ref{fg.cured} we show our estimates of the continuum limit
of $E$ and $P$ in the range $1.3T_c\le T\le4T_c$, where finite volume
effects are small. At $4T_c$ our estimate of the energy density is
about 15\% lower (and the pressure about 19\% lower) than for an ideal
gas.  At this temperature, our estimates of $E$ and $P$ are about 5\%
lower than the earlier determination which used the integral method
\cite{boyd}. While statistically insignificant at present, this
difference is due entirely to the treatment of the region around the
peak of $\Delta$ \footnote{However, note that out use of the 1-loop
expressions for $c_s'$ and $c_t'$, and the consequent inability to make
an optimal choice of scale is the largest uncertainty in this method
for $T\ge1.5T_c$. Such uncertainties can be reduced by appropriate
2-loop computations.}. The speed of sound in the QCD plasma ($c_s$) is
slightly lower than the ideal gas value ($c_s^2=1/3$) at $4T_c$ (but
consistent with it within errors), and falls to
$c_s^2\approx0.1\pm0.025$, at $1.5T_c$.

Our attempt to measure the pressure using perturbatively determined
coefficients may seem mysterious when current practice has converged on
the use of an ``integral'' method \cite{boyd}.  As we have mentioned,
these two methods agree within error bars for $T>2T_c$ but differ near
$T_c$.  The integral method, by construction, gives a continuous
pressure across $T_c$.  However, if true, this is important information
and needs independent verification.  The latent heat density at the
phase transition, $\Delta Q/T_c^4=\Delta S/T_c^3$, where $\Delta S$ is
the jump in the entropy density at $T_c$.  In general this is not the
same as the change in the energy or enthalpy density at the phase
transition. These contain an extra term due to the change in specific
volume at fixed pressure or the change in pressure at fixed volume. In
a quenched lattice simulation, the change in volume is negligible when
changing $T$ by an infinitesimal amount across $T_c$.  As a result, the
pressure should change discontinuously at $T_c$. This is analogous to
boiling water in a closed vessel--- the pressure jumps at the phase
transition. Such a jump in pressure is related to the change in
specific volume, and hence the slope of the phase boundary in the
$(P,T)$ plane through the Clausius-Clapeyron equation. This physics is
missed when the integral method is used to extract the pressure.  This
issue is currently being investigated in a lattice simulation whose
results will be reported elsewhere.

While the observation of scaling on coarse lattices is interesting in
quenched QCD, the real pay off would be in the study of finite
temperature QCD with dynamical light quarks. Since there are inherent
difficulties with simulating QCD with dynamical massless quarks, all
lattice studies have used quarks with mass $m>0$. In such theories,
scaling should be tested at fixed $m/\lms$ (or equivalently, fixed
ratio of $m$ and any hadronic mass scale).

Measurements of $\beta_c$ in QCD with 4 flavours of dynamical staggered
quarks have been performed for $N_t=4$, 6 \cite{nf4a} and 8 \cite{nf4b}.
This last measurement was done with $m/T_c=0.08$. The simulations at
smaller $N_t$ were done at several values of $m$, enabling us to find
$\beta_c$ at $m/T_c=0.08$ by interpolation.  Plaquette values were taken
from a recent finite temperature simulation at fixed $m/T_c$ \cite{our}.
Since it is known that $\Delta_{t,s}/{\cal P}_0$ are less than 0.1\%,
even near $T_c$, we have used $({\cal P}_s+{\cal P}_t)/2$ to determine
$\alphas$. The difference of $\alphas$ measured using ${\cal P}_s$
and ${\cal P}_t$ is taken as an estimate of its error.
We used eq.\ (\ref{scal}) to set the scale.  $T_c/\lms$, when determined
through the bare coupling, changes from 8.2 to 4.7 in going from $N_t=4$
to 8. As shown in Figure \ref{fg.tcms}, when $\alphav$ is used, 2-loop
scaling works much better.

The phase transition in 2-flavour QCD has been studied in greater detail
(see \cite{karsch} for a recent compilation of data). Measurements
of $\beta_c$ using dynamical staggered quarks exist for $N_t\le12$
\cite{betac,plaq}.  Earlier simulations with Wilson quarks (which have
order $a$ lattice artifacts) showed that pion masses were rather high
compared to those obtained with staggered quarks. This problem becomes
less acute on using improved actions for Wilson quarks, and finite
temperature simulations have now been performed with such improved
actions \cite{impwil,cppacs2}.  There have also been some studies with
domain wall Fermions.

We have fixed $\alphav$ and set the scale using published
plaquette values for staggered quarks at several bare quark masses
\cite{plaq,mass}. $T_c/\lms$ for $N_f=2$ shown in Figure \ref{fg.tcms}
are based on the subset of the data which uses staggered quarks
at $m/T_c=0.1$. In contrast to the near constancy of this ratio as
shown in the figure, $T_c/\lms$ computed from the same data using the
bare lattice definition of $g$ falls from 3.9 to 2.8 in going from
$N_t=4$ to 8. To extend this test to other values of $m/T_c$ we have to
interpolate between plaquette values for various quark masses. An upper
limit for the error in this procedure is the actual change
in $\cal P$ between extreme values of the quark masses at
which they are measured. This varies between a few parts in a thousand
at $\beta=5.26$ to about 2\% at $\beta=6$.  This uncertainty in
$\cal P$ translates into a similar magnitude of uncertainty in $T_c/\lms$
and is much smaller than the change when using $6/\beta$ across a
similar range of quark masses.  Excellent scaling of $T_c/\lms$ is seen
also for other values of $m/T_c$.

\begin{figure}[htbp]\begin{center}
   \leavevmode
   \psfig{figure=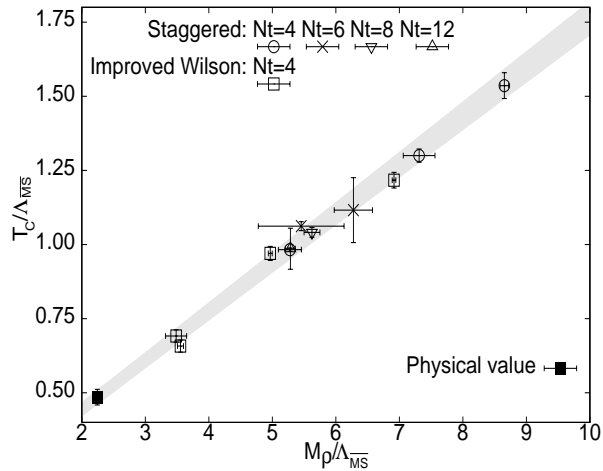,height=6.5cm,width=8cm}
   \end{center}
   \caption{$T_c/\lms$ plotted against $M_\rho/\lms$ and extrapolated
       to the physical value. The band shows the 1--$\sigma$ errors on
       the best fit line to the measurements.}
\label{fg.extrap}\end{figure}

In order to obtain a physically relevant value of $T_c$, it is
necessary to extrapolate $T_c/\lms$ to measured values of the hadron
masses. It would be most interesting to perform this extrapolation in
the quark or pion masses. However, this needs control over the critical
exponents of the theory--- a task we do not attempt here.  Instead we
choose to extrapolate $T_c$ to the physical region in terms of $M_\rho$
\cite{mass}.  There are two reasons for this. First, the ratio
$T_c/M_\rho$ is known to be nearly constant. Secondly, $M_\rho$ is
quite sensitive to finite lattice spacing effects.  The linearity of
the plot of $T_c/\lms$ against $M_\rho/\lms$ in Figure \ref{fg.extrap}
then indicates that, for the chosen data set, finite lattice spacing
effects in $M_\rho$ are under reasonable control.  The value of $\lms$
also needs to be specified.  This depends on how many active flavours
are present at the scale under consideration.  The world average of
$\lms^{(5)}$ (at scales high enough for 5 active flavours) is
$219^{+25}_{-23}$ \cite{pdg}. At lower scales, with only three active
flavours, $\lms^{(3)} = 343^{+31}_{-28}$, using the prescription of
\cite{pdg} to match across flavour thresholds.

The physical value of $T_c$ is obtained by extrapolating to the real-world
value of $M_\rho/\lms^{(3)}$. This is done by fitting a straight line to
the data including staggered and improved Wilson quarks \cite{impwil}. We
have taken the two sets of data together since they both have cutoff
effects of order $a^2$.  The extrapolation gives
\beq
   T_c/\lms=0.487\pm0.023, \qquad{\rm and}\qquad
   T_c = 167 \pm 9 {}^{+15}_{-14}\qquad\qquad(N_f=2{\rm\ QCD}).
\label{tc}\eeq
The error in the ratio above reflects only the statistical errors in
various measurements.  The first error in $T_c$ is purely from
extrapolation and the second set from the errors on $\lms^{(3)}$.
Since they come from independent sources, it is possible to add them in
quadrature. This result is consistent with a recent estimate using
$O(4)$ critical indices to scale $T_c$ as the pion mass is taken to its
physical value \cite{cppacs2}. However, there are possibly large
systematic uncertainties.  Currently the least well-understood problem
is whether extrapolating one hadron mass to its measured value also
takes all other hadron masses to their correct values.  Until this
issue is settled, all estimates of $T_c$ must be considered
preliminary.

\begin{figure}[htbp]\begin{center}
   \leavevmode
   \psfig{figure=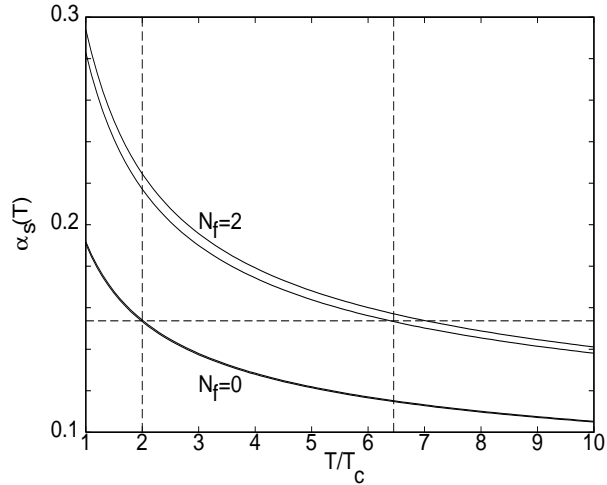,height=6.5cm,width=8cm}
   \end{center}
   \caption{$\alphas$ at 2-loop order in the $N_f=0$ and $N_f=2$
       theories at the scale optimised for the three-dimensional gauge
       coupling in dimensionally reduced theories, shown as a function
       of the temperature.  The bands arise from the quoted errors in
       the measurement of $T_c/\lms$.  This plot is used to limit the
       validity of dimensional reduction for $N_f=2$ as explained in
       the text.}
\label{fg.dimred}\end{figure}

We end with a small application of our measurements of $T_c/\lms$.
This addresses the question of the temperature range where dimensional
reduction (DR) is expected to be valid.  Perturbative matching of the
$T>0$ four dimensional theory with DR fixes the effective couplings in
the latter \cite{keijo}. There is good numerical evidence that the
$SU(3)$ pure gauge theory agrees with its DR version for $T>2T_c$
\cite{saumen}. If the value of the strong coupling were the determining
factor in the agreement, then this would imply that for $N_f=0$,
$\alphas$ is small enough at $2T_c$ to trust this expansion. If for
$N_f=2$ no other physics becomes more important, then the matching
should be equally reliable when the coupling is as small. Now we ask
for the temperature at which $\alphas$ in the $N_f=2$ theory takes on
the value that it had at $2T_c$ for $N_f=0$.  Using the values of
$T_c/\lms$ already extracted, we are in a position to do this.  The
optimal scale choice for $\alphas$ has been investigated in
\cite{keijo}. This depends, of course, on $N_f$ and the quantity for
which the perturbation theory is optimised. Choosing the scale which
optimises the three-dimensional gauge coupling we find the dependence
of $\alphas$ on $T$ shown in Figure \ref{fg.dimred}.  This tells us
that for $N_f=2$ DR can be used only for $T>6.5T_c$. Using scales
optimised for other quantities, and varying the computation of the
strong coupling from 2-loop to 3-loop order, we find that the lower
limit of the range of validity of DR for $N_f=2$ varies between $6T_c$
and $8T_c$.

In summary, we demonstrated that the lattice data on strong interaction
thermodynamics obey QCD scaling relations very well and allow some
continuum physics to be extracted on fairly coarse lattices. $T_c$ can
be measured with precision of about 4\%, taking into account both
statistical errors as well as systematic uncertaintites such as
renormalisation scheme dependence.  Simulations including dynamical
quarks, when extrapolated to physical values of the $\rho$ meson mass
give statistical errors of about 4\% in $T_c/\lms$.  Renormalised
perturbation theory at 1-loop order seems to be able to give a good
account of the pressure, energy density and the speed of sound for
$T>1.5T_c$, leading to hopes that improved actions and higher loop
orders can yield continuum physics at small expense.  A study of finite
volume effects on thermodynamics closer to $T_c$ will be reported
elsewhere.

It is a pleasure to thank Urs Heller for making available the full results
of the fits reported in \cite{heller}, and Rajiv Gavai for many discussions.

 \end{document}